\documentclass[10pt,final,twocolumn,twoside]{IEEEtran}
\usepackage{framed}
\usepackage{graphicx} \usepackage{amsmath} \usepackage{amssymb}
\usepackage{color}
\usepackage{amssymb}
\usepackage{amsthm} 
\usepackage{multirow}
\usepackage[linesnumbered,ruled,vlined]{algorithm2e}
\SetKwComment{Comment}{$\triangleright$\ }{}
\newlength\algowd

\usepackage{cite} 
\usepackage[latin1]{inputenc}
\usepackage[caption=false,font=footnotesize]{subfig} 
\usepackage{stfloats}
\usepackage{url}
\usepackage{mathtools}
\usepackage{varwidth}

\usepackage{xpatch}
\makeatletter
\patchcmd\@makecaption{\\}{.~}{}{\fail}
\makeatletter

\newcommand{\x}[1]{\mathbf{x}^{<#1>}}
\newcommand{\y}[1]{\mathbf{y}^{<#1>}}

\newcommand{\mbf}[1]{\mathbf{#1}}     \newcommand{\mbs}[1]{\boldsymbol{#1}}

\newcommand{\what}[1]{\widehat{#1}}

\begin{document}

\title{Proactive Mobility Management of UEs using Sequence-to-Sequence Modeling}
\date{\today}
\author{Vijaya~Yajnanarayana \\
 Ericsson Research\\
 Email: vijaya.yajnanarayana@ericsson.com}
\maketitle
\thispagestyle{empty}
\pagestyle{empty}

\begin{abstract}
  Beyond 5G networks will operate at high frequencies with wide bandwidths. This brings both opportunities and challenges. Opportunities include high throughput connectivity with low latency. However, one of the main challenges in these networks is due to the high path loss at operating frequencies, which requires network to be deployed densely to provide coverage. Since these cells have small inter-site-distance (ISD), the dwell-time of the UEs in these cells are small, thus supporting mobility in these types of dense networks is a challenge and require frequent beam or cell reassignments. A pro-active mobility management scheme which exploits the historical trajectories can provide better prediction of  cells and beams  as UEs move in the coverage area. We propose an AI based method using sequence-to-sequence modeling for the estimation of handover cells/beams along with dwell-time using the trajectory information of the UE. Results indicate that for a dense deployment, an accuracy of more than $90$ percent can be achieved for handover cell estimation and very low mean absolute error (MAE) for dwell-time.

\end{abstract}

\section{Introduction}
\label{sec:intro}
Future cellular networks such as 6G will operate in high frequencies with wide bandwidth. Due to the high path loss in these frequencies, the network needs to be densely deployed  to provide coverage,  this leads to small inter-site distance (ISD). The small ISD deployment coupled with high mobility use-cases of future networks results in small dwell-time in cells/beams requiring frequent handovers (HO) and beam switches. This leads to frequent measurements from UEs causing excessive battery drain. These measurements are transferred over control channels resulting in excessive control-data flow. Also, the process of performing measurements and radio resource control (RRC) procedures to transfer them introduce latency causing interruptions in the data flows.  Many existing methods predict only next potential base-station. In a densely deployed network, a typical call can have multiple handovers. In such scenarios, knowledge about not only the next handover BS, but also the next few potential BS\footnote{Basestation and cell are interchangeably used} handovers in a sequential order, together with the  dwell-times in those BSs can aid in provisioning, load-balancing, etc. As the cellular communication system starts to evolve into higher frequency bands, the spatial resolution of the beams used to serve UEs will be high. This results in frequent serving beam switches or inter-beam handovers within a serving cell.

\begin{figure}[t]
  \centering
  \fbox{\includegraphics[width=3 in]{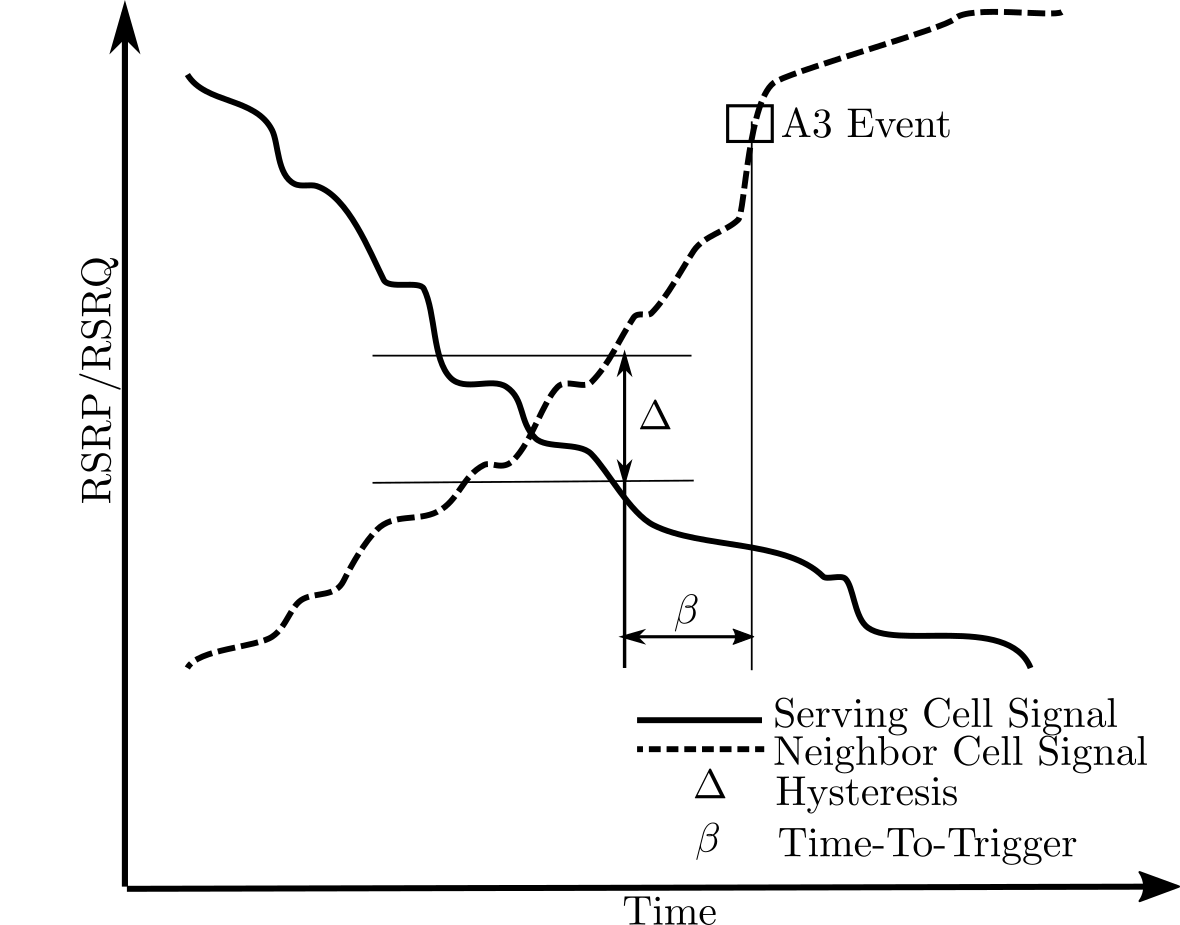}}
  \caption{Illustration of the HO mechanism in cellular networks}
  \label{fig:handover}
\end{figure}

The current intra-frequency HO procedures  described in \cite{3gpp.36.331},\cite{handover-book} are reactive in its approach. Typically, when UE enters a particular RRC event it will perform neighbor cell measurements and report them to the BS.  The UE is said to have entered a particular RRC event when a corresponding entering condition is satisfied. These conditions are signaled by the serving BS in the form of parameters such as thresholds, offset, and hysteresis.  Due to the frequent handovers in densely deployed future networks, the UEs will enter these events frequently, resulting in more frequent measurements causing excessive battery consumption and increased latency. These measurements are transferred over the control channel leading to the excessive uplink control data flow. In \cite{3gpp.36.331}, \cite{3gpp.36.133}, the RRC procedures and protocols involved in mobility are discussed including the RRC signaling mechanism for configuring the UE to event triggers. Fig.~\ref{fig:handover} illustrates an A3 event based handover mechanism where the event criteria is met when the access-beam power of the neighbor cell is  greater than a threshold denoted by hysteresis value ($\Delta$) and holds for a duration greater than time-to-trigger ($\beta$) (refer to Fig.~\ref{fig:handover}). 

Similar to the intra-frequency cellular HO, the beam switching (inter-beam HO) process requires a complex beam management process which typically involve beam-sweeping, performing measurements on the beams, determining the suitable beam(s) and reporting it to the BS. This process introduces lot of overhead in terms of computation, delay and battery consumption.

In this paper, we propose a sequence-to-sequence modeling approach for the intra-frequency cellular HO and beam switching (inter-beam HO) problems  discussed above, wherein an AI agent is trained using the  sequential evolution in the radio parameter space  by exploiting the frequently employed mobility patterns to predict the future HO cells and beams without the need for measurements.

\subsection{Related Work}
\label{ss:rw}
There exists several prior works where the handover optimization is achieved by  optimizing parameters  such as time-to-trigger ($\beta$) and hysteresis ($\Delta$) based on the deployment environment.  For example, in \cite{Leu-1} and \cite{Leu-2} authors propose methods where the parameter optimization is achieved to reduce ping-pong. In \cite{Wu}, the optimization of HO between macro and femto BS is achieved by exploiting the UE information such as velocity, received signal strength indication (RSSI), etc. Similar to our work, machine learning has been applied in several HO optimization problems. In a typical  hybrid cellular networks supporting both unmanned aerial vehicles (UAVs) and terrestrial UEs, the  UAVs are served by the side-lobes of the BS antenna thus requiring frequent HOs \cite{Drone1}. In \cite{DroneMobility}, the author's propose an reinforcement learning (RL) based HO approach for UAVs. The proposed model learns the fragmented 3D-coverage described in \cite{Drone1}, while trading off throughput, ping-pong and radio link failures (RLF).  In \cite{VJ-rlpaper}, authors propose a method where neighbor cell measurements are captured by the serving cell in a periodic reporting mode and  measurements are sent to an edge server to predict the optimal HO cell using RL. In \cite{VJ-platoon-patent}, authors propose a novel scheme for mobility in platooning, where  measurements and RRC signaling from the individual UEs of the platoon are avoided using the measurements collected from platoon head. In \cite{memon-2019-fogcomputing}, a method for internet of vehicles (IoV) to predict the fog-node not only based on position but also on the parameters such as compute and memory availability  is proposed. In \cite{VJ-RRC-patent}, a sub-cell level resolution for mobility is proposed, where the trajectories of UEs are mined using an unsupervised learner to automatize the RRC actions.

\subsection{Contribution}
In contrast to our work, in all the above discussed works, the proposed methods are agnostic to the historical information such as:
\begin{itemize}
\item The various cells the UE has traversed together with the time spent in them (known as dwell-time) for intra-frequency cellular HO problem
\item Historical evolution of the UE in the beam-space of serving BS for beam reassignment or inter-beam HO problem.
\end{itemize}
In this paper, this historical information is exploited by AI methods to aid in better mobility management in wireless systems. The main contributions of the paper are listed below:
\begin{itemize}
\item A sequence-to-sequence modeling approach for the intra-frequency cellular HO and beam switching problems
\item An estimation method to predict the sequence of future HO BSs that the UE may handover to and the amount of time it may spend in those BSs
\item An estimation method to predict the sequence of future beams UE may switch to within the serving cell
\item Performance evaluations of the proposed estimation methods using both synthetic data from a simulator and live mmWave network data 
\end{itemize}
All the proposed estimation algorithms are AI based and does not require any measurements by the UE.

\section{System Overview}
\label{sec:overview}
In intra-frequency HO problems, as UE moves in the network area, it accumulates historical information such as last visited cells, dwell-time in those cells, etc. This information is updated when UE perform cell-selection/reselection procedures in RRC\_IDLE state or HO procedures in RRC\_CONNECTED state. This can be queried by the serving BS using RRC signaling \cite{3gpp.36.331} which can then be used to construct the sequence of cell-ID and dwell-time evolution needed to train the AI agents for predicting the future cell-ID and dwell-times. 

The trajectory of the UE within a cell can be represented as evolution in serving BS's beam-space. Here the serving-beam together with the neighbor-beam measurements form a hyper-space in which the trajectory of the UE is evolving. This sequential evolution can be exploited to train the AI agents for predicting future beam switches.

A crowd sourced approach can be employed to capture   sequences of evolution for different  UEs having different mobility patterns using traditional methods based on measurements. From these sequences an AI agent can be trained to predict future beam-ID  or  cell-ID / dwell-time without any measurements from the UE.

\begin{figure}[t]
  \centering
  \fbox{\includegraphics[width=2.5 in]{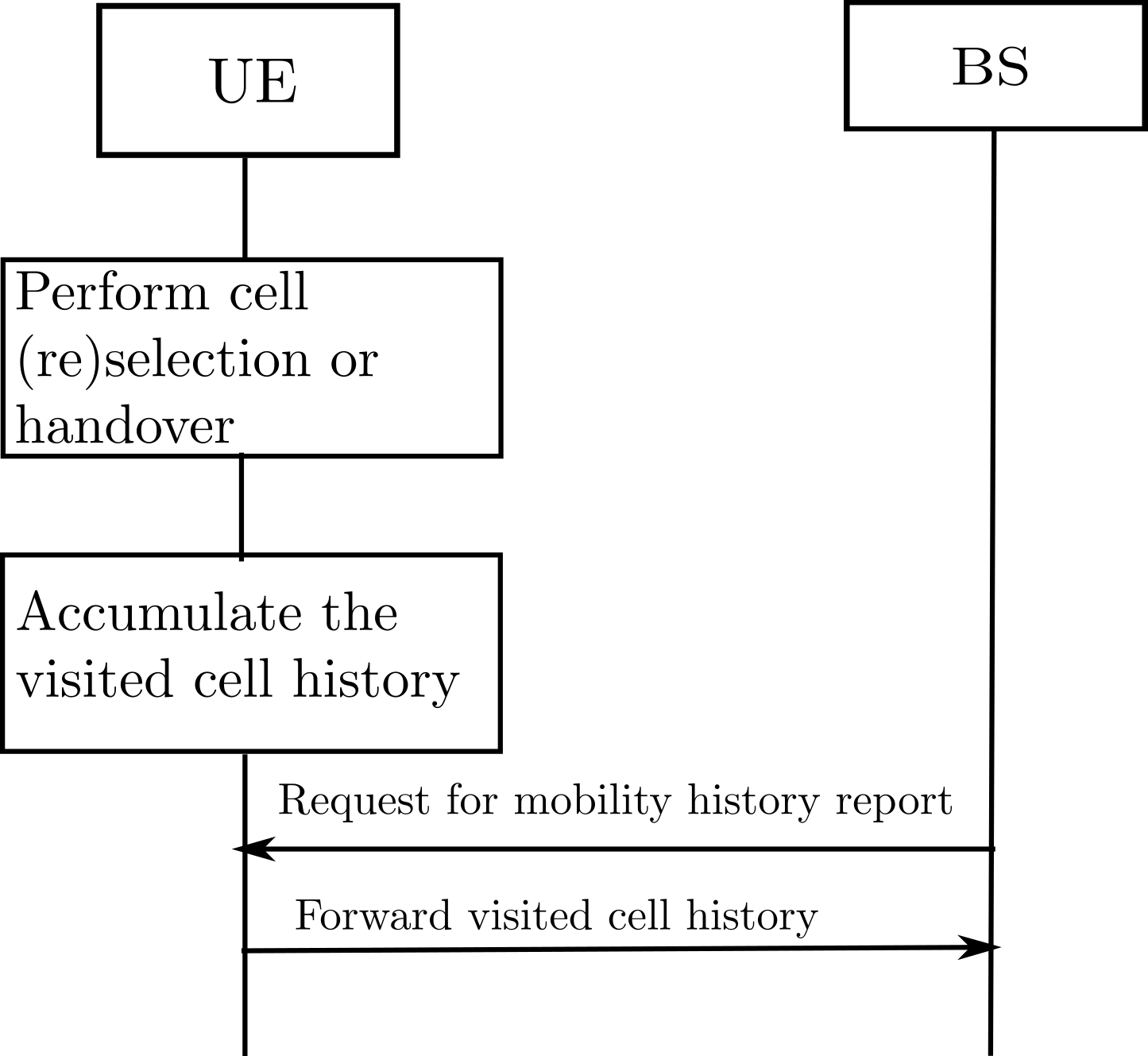}}
  \caption{Mobility history can be obtained in the BS using RRC signaling}
  \label{fig:signalling}
\end{figure}
\begin{figure*}[t]
  \centering
  \includegraphics[width=6 in]{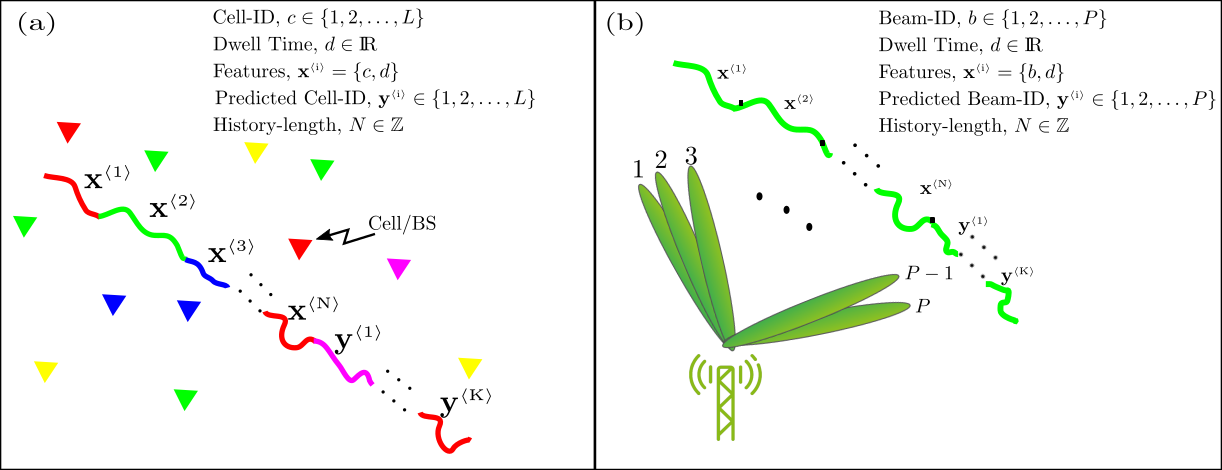}
  \caption{UEs trajectory is defined as evolution of radio parameter $\mbf{x}$, over time. In (a) cell-IDs together with dwell-time is used for prediction while in (b) serving-beamIDs are used. The trajectory estimation can be posed as a sequence-to-sequence estimation where we are interested in predicting the HO cells in (a) while in (b) we are interested predicting the future serving beam-IDs.}
  \label{fig:UE_mobility}
\end{figure*}
\section{Method}
\label{sec:method}
The motion of a UE can be represented as evolution of  radio parameters in time.  For intra-frequency HO problems this could be  cell-ID ($c$), dwell-time ($d$) and for beam switching problem this could be beam-ID ($b$).  This is further illustrated in Fig.~\ref{fig:UE_mobility}. The $\x{i}$, denotes  hyper-space of radio parameters values. The trajectory of the UE can be crudely expressed as an evolution of $\mbf{x}$ in time. The UE trajectory prediction can be posed as sequence-to-sequence prediction problem, where after observing sequence $\x{1},\ldots,\x{N}$, we are interested in predicting sequence $\y{1}, \ldots,\y{K}$. This is further illustrated in Fig.~\ref{fig:UE_mobility}, where $N$ is the history length and $K$ is the number of future steps to be predicted.

The optimal solution for this problem, is to maximize the $\mbox{a-posteriori}$ probability with the choice $\y{i},i\in\{1,2,\ldots,K\}$, that is
\begin{equation}
  \label{eq:rnn}
  \begin{split}
  &\what{\mbf{y}}^{<1>},\ldots,\what{\mbf{y}}^{<K>} = \\
  &\quad \underset{\mbf{y}^{<1>}, \ldots ,\mbf{y}^{<K>}}{\mbox{argmax}}{p\{\mbs{x}^{<1>}, \ldots,\mbs{x}^{<N>}|\mbf{y}^{<1>}, \ldots, \mbf{y}^{<K>}\}} .
  \end{split}
\end{equation}
We propose an AI based learning using recursive neural network (RNN) to solve for the above optimization problem. The structure is as shown in the Fig.~\ref{fig:RNN}.

\section{Simulations}
\label{sec:sim}
\begin{figure}
  \centering
  \includegraphics[width=2.5 in]{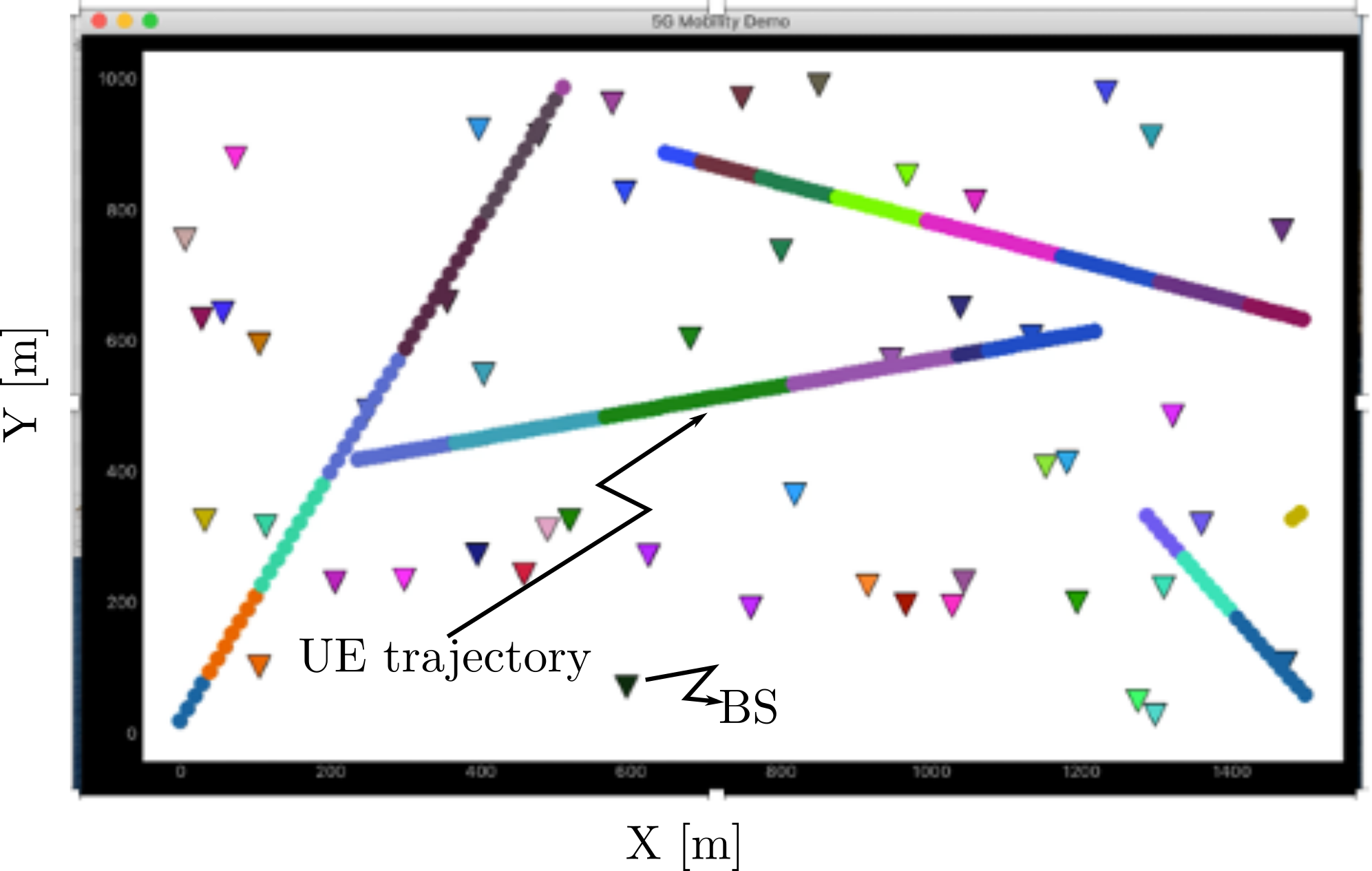}
  \caption{A demonstration of a mobility pattern created using a simulator. The UEs move in straight lines with random slopes.}
  \label{fig:deployment}
\end{figure}

For the intra-frequency HO problem as shown in the Fig.~\ref{fig:UE_mobility}(a),  we created different mobility patterns in a 2D-map having 50 BS nodes randomly dropped in a $1\mbox{ km}^2$ area to assess the performance.  The BSs are configured to have $43~\mbox{dBm}$ of power with $3$ sector antennas. We consider a  line-of-sight (LOS) path-loss model having a path loss exponent of $3.1$. The UEs are made to move in straight lines with different slopes, when the UE hits the end of the 2D-raster, it will randomly relocate and move with a random slope. A screen capture of the mobility demo GUI is shown in  Fig.~\ref{fig:deployment} which illustrates this further. 

\begin{figure}[b]
  \centering
  \fbox{\includegraphics[width=3 in]{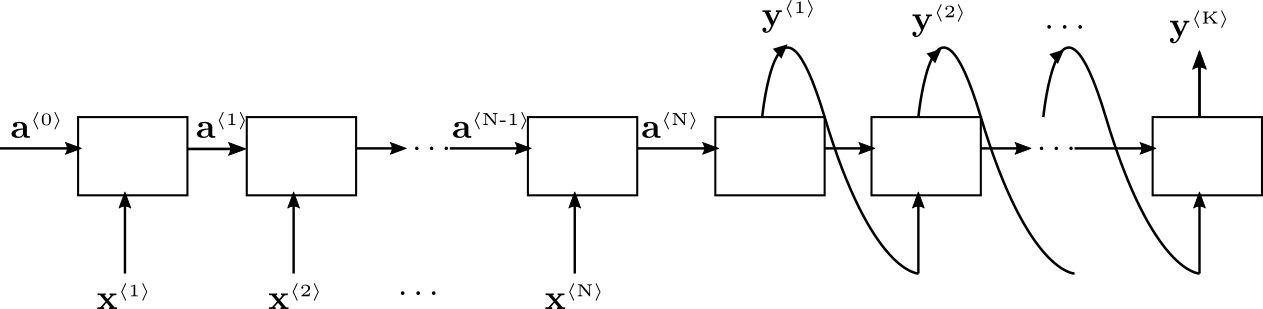}}
  \caption{RNN structure for sequence-to-sequence estimation of UE Trajectory using historical information}
  \label{fig:RNN}
\end{figure}

In the following, we describe how the  training and validation data-set are built for assessing the performance of the RNN structure shown in the Fig.~\ref{fig:RNN} for intra-frequency HO problem. We consider the deployment as shown in the Fig.~\ref{fig:deployment}. The triangles in the Fig.~\ref{fig:deployment} depicts the BSs. The different colors in the segments of the straight lines shows the associated serving basestation.  The UEs are configured in periodic reporting mode, at each step an HO inference is made, if an HO is initiated the mobility history is updated. As the UEs move along the chosen mobility pattern, the UEs will make handover to the BS with best synchronization beam. The color of the segments  in Fig.~\ref{fig:deployment}, indicate the color of the serving BS. The sequences of the traversed basestations together with the dwell-time are used to build the training and validation sets which in turn  is used to train the RNN structure shown in the Fig.~\ref{fig:RNN}.

\begin{figure}[t]
  \centering
  \fbox{\includegraphics[width=3 in]{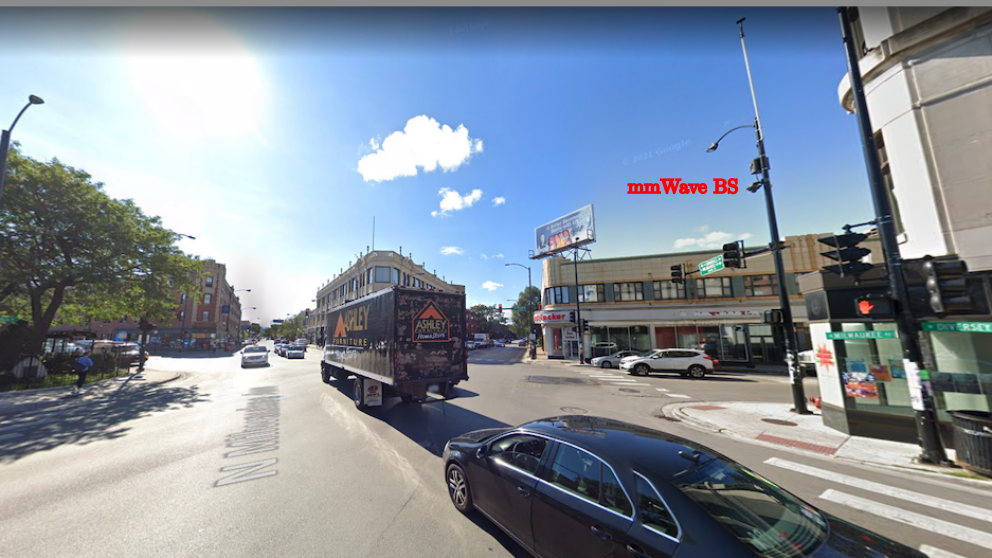}}
  \caption{The road-side mmWave BS mounted on the traffic light pole is used to capture the beam-ID  traces used for training the RNN structure shown in Fig.~\ref{fig:RNN}. }
  \label{fig:DataForBeamID}
\end{figure}

\begin{figure}[t]
 \centering
  \fbox{\includegraphics[width=2.5 in]{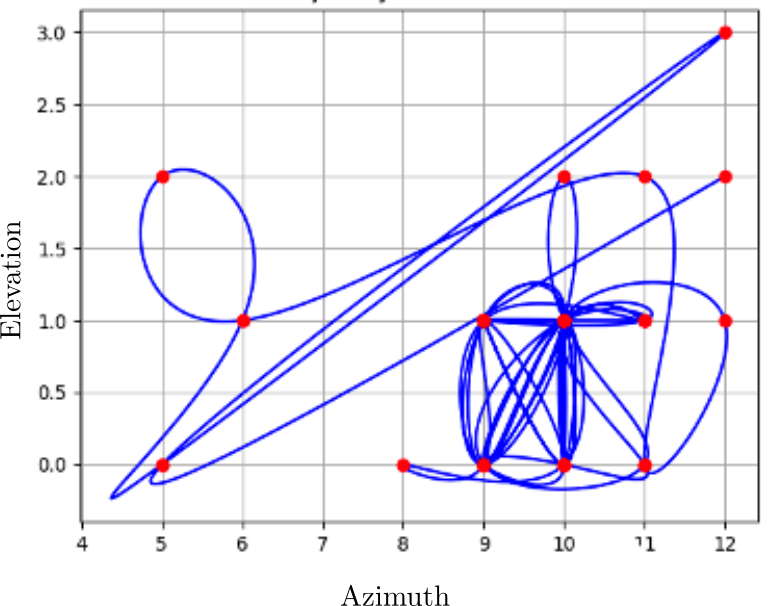}}
  \caption{Trajectory of UE in beam space after considering the azimuth  and elevations of the beams.}
  \label{fig:TrajectoryInBeams}
\end{figure}

For the beam switching problem shown in Fig.~\ref{fig:UE_mobility}(b), we used the data captured from a live  5G mmWave network operated by Ericsson in an urban area of Chicago. The setup contained a road side mmWave basestation mounted on a traffic light pole as shown in Fig.~\ref{fig:DataForBeamID} with number of beams, $P=68$. As UEs travel along the road, the sequence of various beam transitions are captured. Fig.~\ref{fig:TrajectoryInBeams} shows the trajectory (sequence of beam-ID transitions) of a UE after considering the azimuth  and elevations for the selected beam-IDs. These trajectories are then used to train the RNN structure shown in Fig.~\ref{fig:RNN} for the beam switching problem.

\begin{table}[h]
    \caption{Summary of the simulation parameters. $L$ indicates number of features considered.}
  \centering
  \begin{tabular}{ p{1.05in}|p{1.05in}|p{1.05in}}
    \hline 
    \multicolumn{2} {c} {Parameter}  &  Value \\
    \hline
    \multirow{2}{10em}{Neural Net Structure} &  Input Layer & $(N, L)$ \\
                                     & RNN Layer & 100 RNN units \\
                                     & Output Layer & Dense layer\\
                                     & Activation & ReLU for RNN and sigmoid for dense layer \\ 
    \hline
    \multirow{2}{2em}{Loss} & BS prediction & Categorical cross entropy (CCE)\\
                                     & Dwell-time estimation & Mean absolute error (MAE) \\
    \hline
    Optimizer & & Adam \\
    \hline
  \end{tabular}
  \label{tab:nn-param}
\end{table}
\section{Results}
\label{sec:results}
In this section, we assess the performance of the proposed method under various conditions.
\subsection{Using historical cell-IDs  for predicting the next sequence of cell-IDs for handover}

\begin{figure}[t]
  \centering
  \fbox{\includegraphics[width=2.5 in]{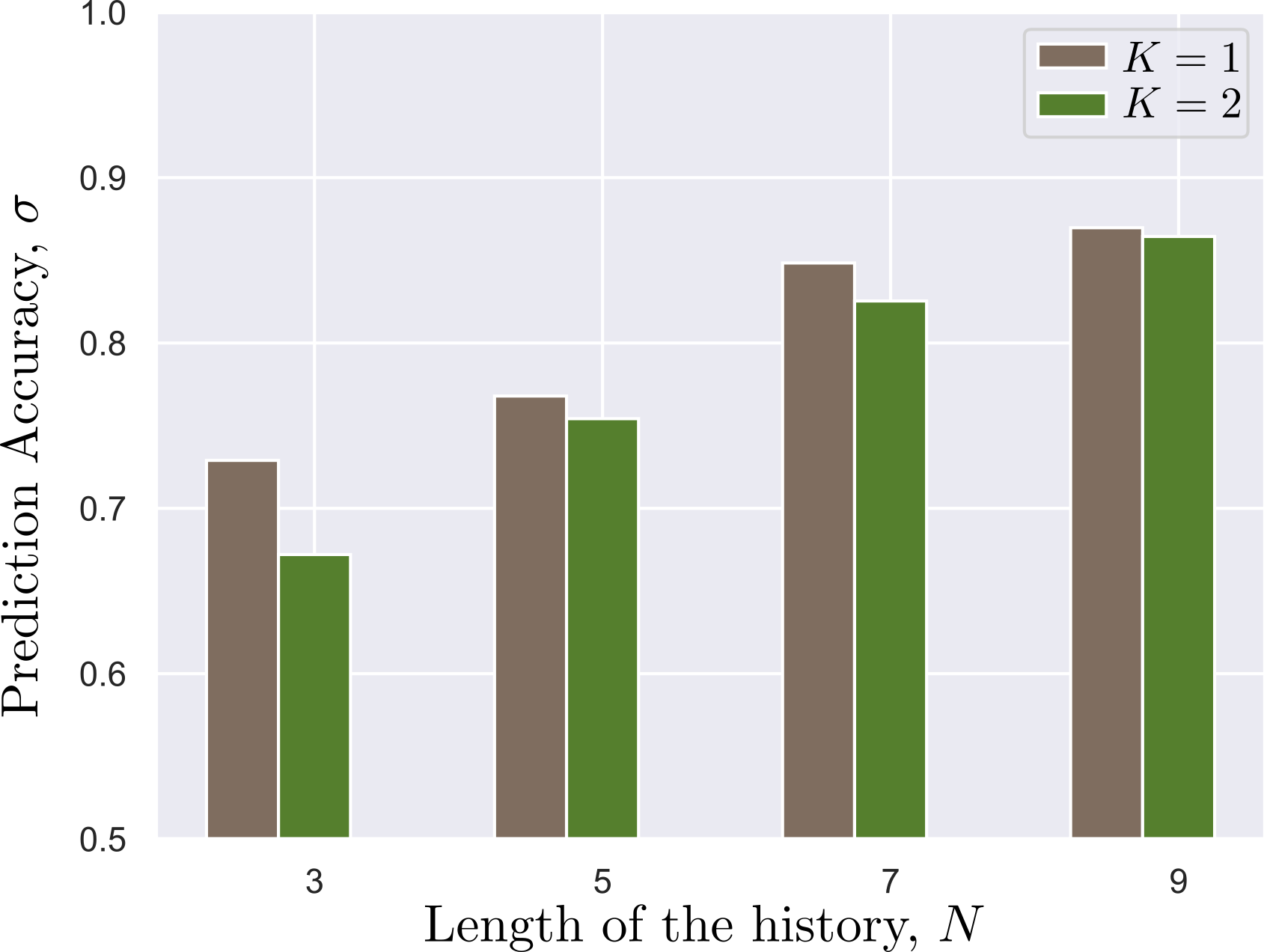}}
  \caption{The accuracy in predicting the sequence of handovers in future, with different mobility history lengths (N varying from $3$ to $13$). The accuracy is poor when predicting farther into the future (large $K$) with less history (small $N$)}
  \label{fig:rx-1}
\end{figure}

\begin{figure}[t]
  \centering
  \fbox{\includegraphics[width=2.5 in]{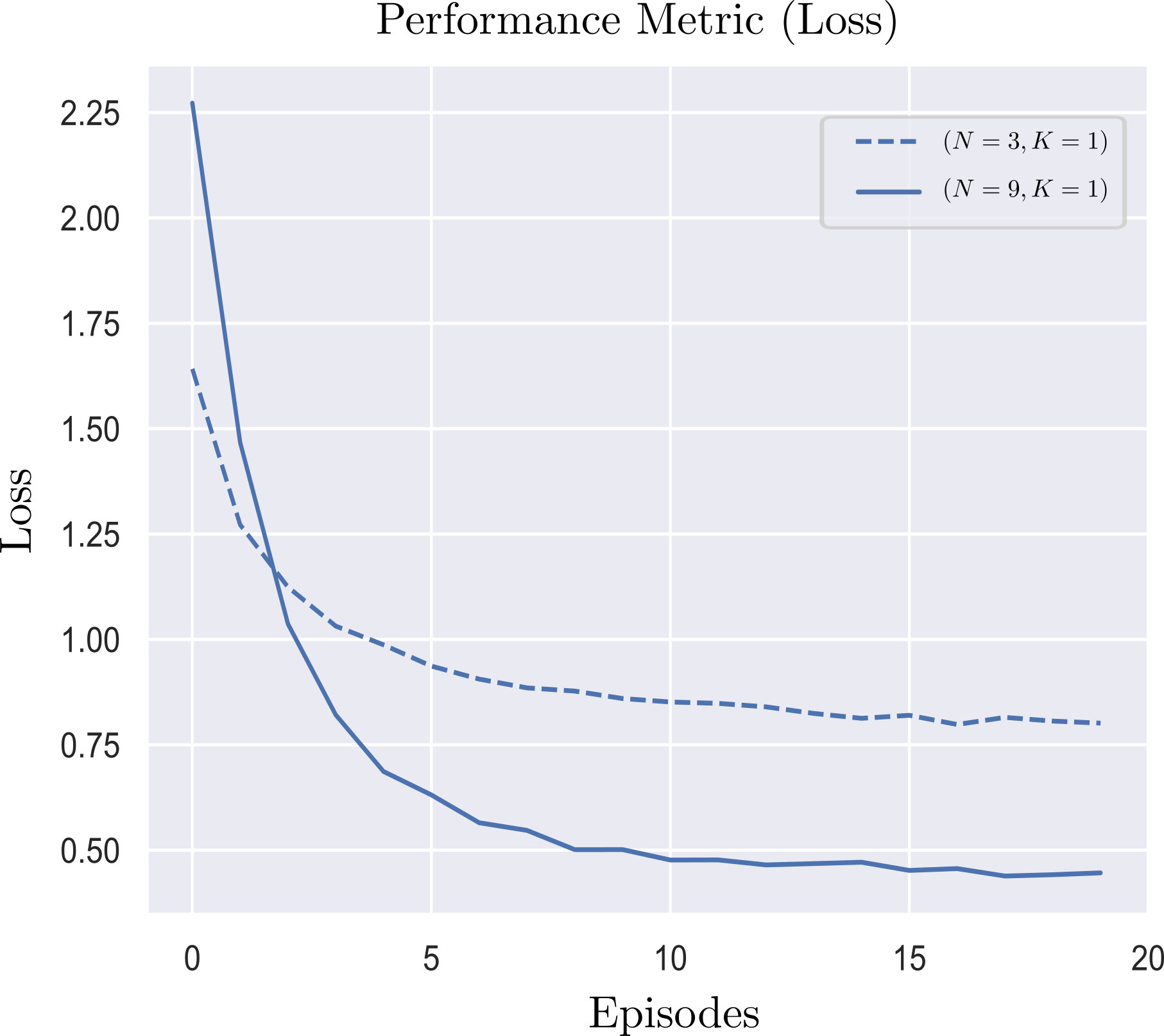}}
  \caption{Loss verses episodes plot indicating the convergence of the method within $10$ episodes}
  \label{fig:rx-2}
\end{figure}

We assess the performance for a simple scenario considering only the BSs (cell-IDs) that the UE has handovered to before reaching the current BS. Our objective is to predict the sequence of BSs the UE will HO to in future, therefore we have
\begin{equation}
  \label{eq:rx-1-A}
  \x{i}=\{c_i\} , c_i\in\{1,\ldots,L\},  i\in\{1,\ldots,N\},
\end{equation}
and
\begin{equation}
  \label{eq:rx-1-B}
  \y{j}=\{c_j\} ,  c_j\in\{1,\ldots,L\}, j\in\{1,\ldots,K\},  
\end{equation}
where $c_i$ denotes the cell-ID of the $i$-th element in the mobility history information (MHI) sequence.  The proposed structure in Fig.~\ref{fig:RNN} is configured using the configuration shown in the Table~\ref{tab:nn-param}. Fig.~\ref{fig:rx-1} shows the performance of the proposed method. The accuracy of the prediction increases with the historical length of the sequence  ($N$) and saturates due to the decorrelation of the prediction with the cell-ID which are far back in the sequence. Also, comparing  the accuracy of the prediction  for $K=1$  and $K=2$, the accuracy falls while predicting longer handover sequences (farther into the future)  with same amount of historical information (i.e., with fixed $N$). Fig.~\ref{fig:rx-2} shows the categorical cross entropy (CCE) loss  verses episodes indicating that the  algorithms converges within $10$ episodes. 

\subsection{Using historical cell-IDs  for predicting the dwell-time in the next Cell}
\begin{figure}[t]
  \centering
  \fbox{\includegraphics[width=2.5 in]{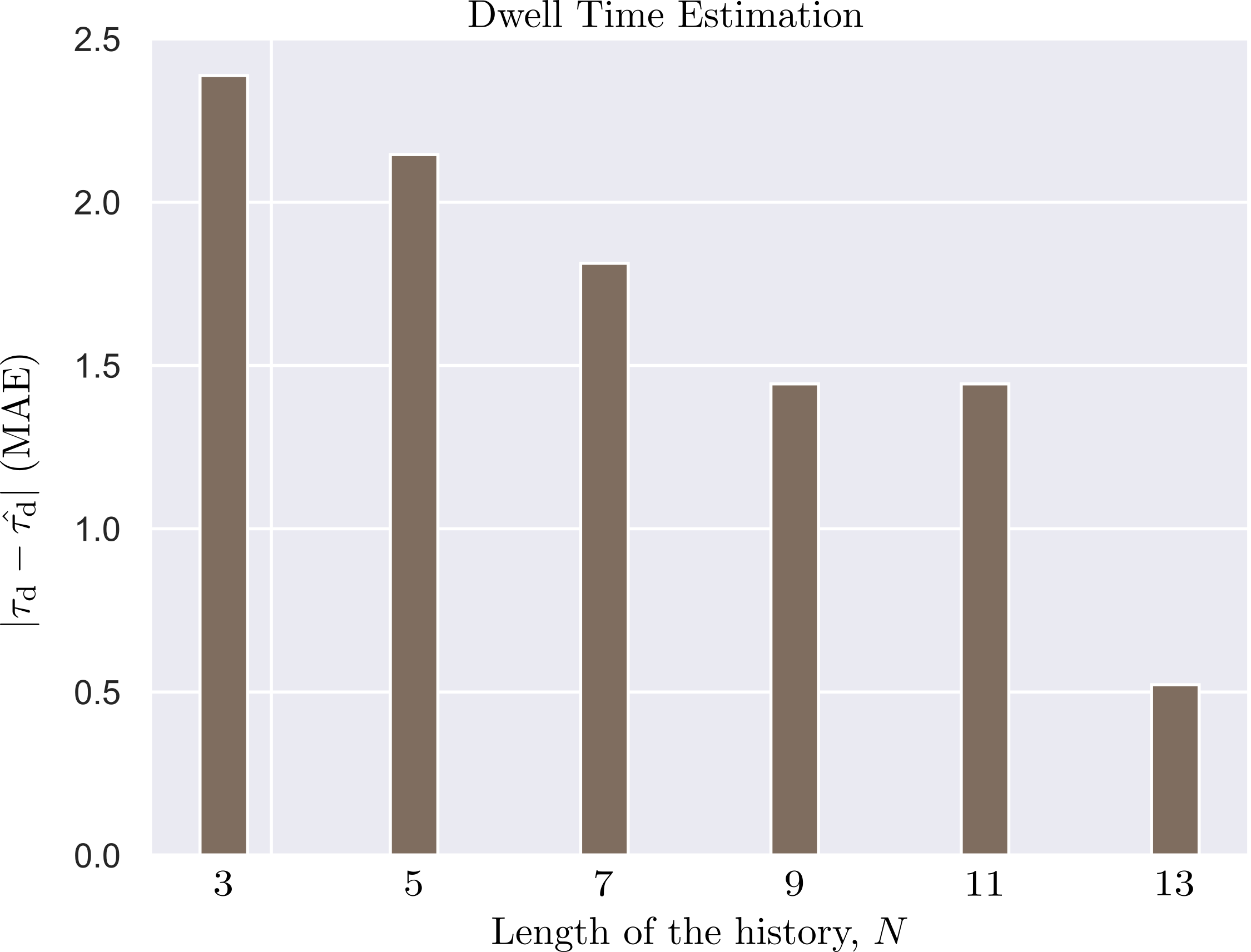}}
  \caption{MAE performance with different mobility history lengths}
  \label{fig:ry-1}
\end{figure}

\begin{figure}[t]
  \centering
  \fbox{\includegraphics[width=2.5 in]{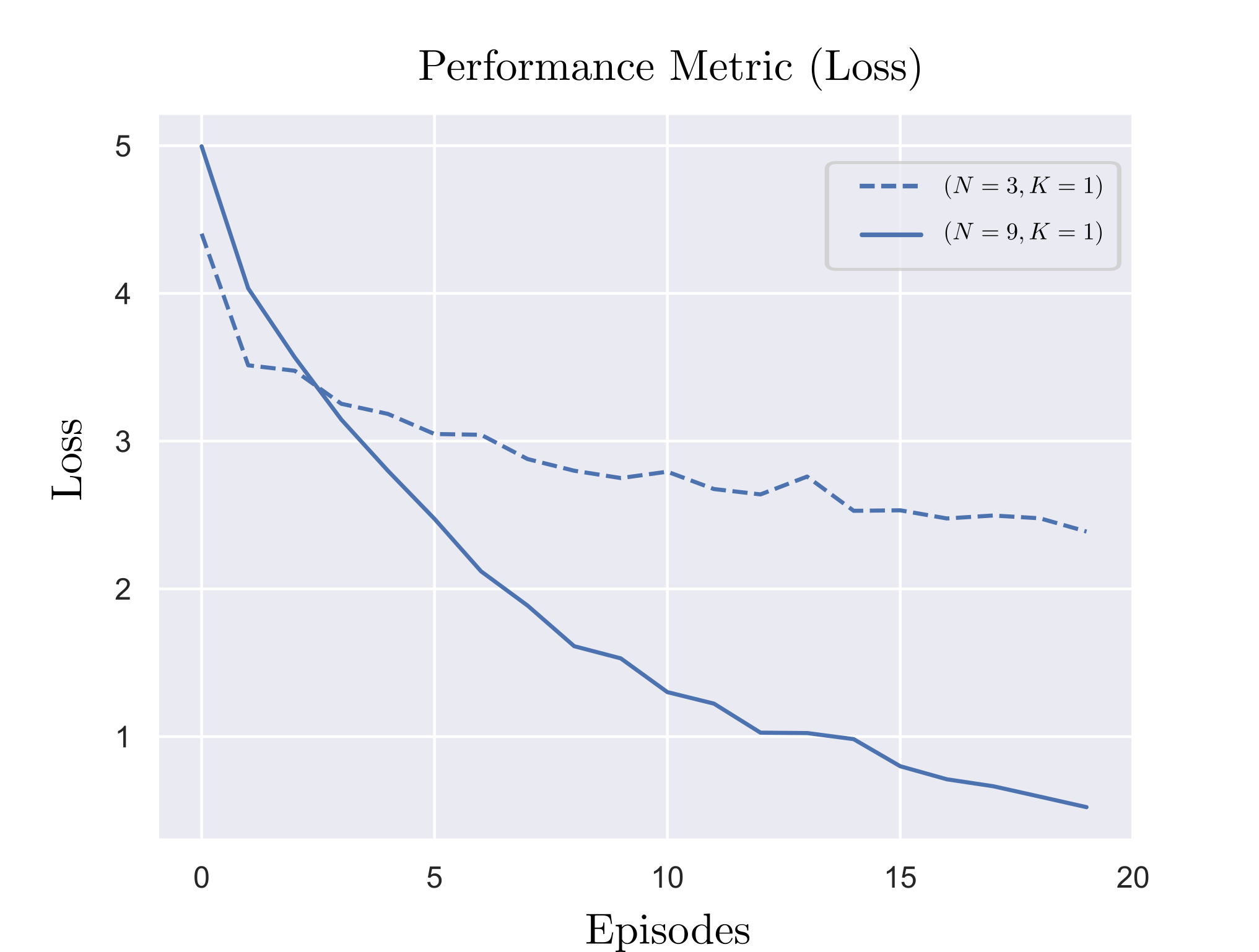}}
  \caption{The loss verses episodes}
  \label{fig:ry-2}
\end{figure}

The objective of this formulation is to exploit the structure shown in the Fig.~\ref{fig:RNN} to estimate the dwell-time of the UE in the next HO cell. The dwell-time is measured in terms of number of reporting steps in the periodic reporting mode. The feature and prediction variables are as given in the equation below:
\begin{align}
  \begin{split}
  \label{eq:ry-1-A}
  \x{i}&=\{c_i,d_i\} , c_i\in \{1,\ldots,L\}, \\
         & \quad \quad i\in\{1,\ldots,N\},  d_i\in {\rm I\!R},
         \end{split}
\end{align}
and
\begin{equation}
  \label{eq:ry-1-B}
  \y{j}=\{d_j\} ,  d_j\in{\rm I\!R}, j\in\{1,\ldots,K\},  
\end{equation}
where $d_i$ denotes the dwell-time at the $i$-th BS. The proposed structure in Fig.~\ref{fig:RNN} is configured for simulation with the parameter shown in the Table~\ref{tab:nn-param}. Since we are interested in the dwell-time in the next BS, we set $K=1$. Fig.~\ref{fig:ry-1} shows the performance of the proposed method in terms of mean-absolute-error (MAE). Fig.~\ref{fig:ry-2} shows the loss performance across the episodes. Note the reduction in MAE with the increased history information ($N$). 

\subsection{Multi-step handover estimation using joint information of historical cell-IDs and dwell-times}
In this section, our objective is to utilize historical cell-ID with associated dwell-time in them to estimate a multi-step prediction of future handover BSs. So in the structure shown in Fig.~\ref{fig:RNN}, we employ the features and prediction variables as given in the equations below:
\begin{align}
  \label{eq:rz-1-A}
  \begin{split}
    \x{i}&=\{c_i,d_i\} , c_i\in \{1,\ldots,L\}, \\
    &\quad i\in\{1,\ldots,N\},  d_i\in{\rm I\!R},
    \end{split}
\end{align}
and
\begin{equation}
  \label{eq:rz-1-B}
  \y{j}=\{c_j\} ,  j\in\{1,\ldots,K\}.  
\end{equation}

\begin{figure}[t]
  \centering
  \fbox{\includegraphics[width=2.5 in]{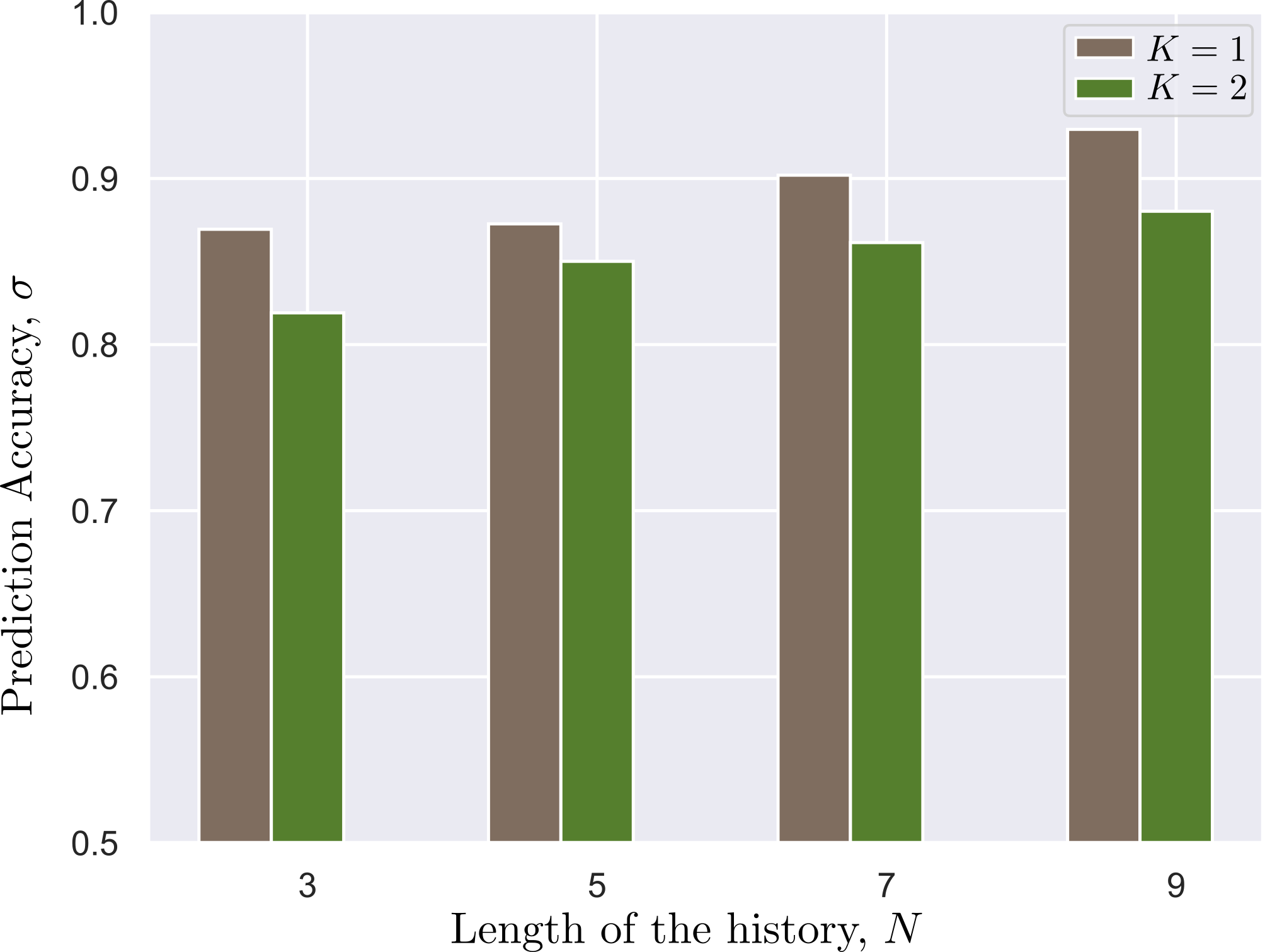}}
  \caption{Single ($K=1$) and Multi ($K=2$) step prediction performance with a multi-variate RNN with history consisting of  dwell-time and cell-IDs }
  \label{fig:rz-1}
\end{figure}

The proposed structure in Fig.~\ref{fig:RNN} is configured for simulation using parameters  of Table~\ref{tab:nn-param}. The performances by setting $K=1$ and $K=2$ is shown in Fig.~\ref{fig:rz-1}. Notice that with increased $K$ there is a slight loss of performance as we are trying to predict far into the future with same amount of historical information. By comparing the accuracy  in Fig.~\ref{fig:rz-1}  with  Fig.~\ref{fig:rx-1} discussed in Section~\ref{sec:sim}-A, notice an improved accuracy by converting the structure into a multivariate RNN formulation involving both historical cell-IDs and associated  dwell-times.

\subsection{Serving beam  prediction using traces from a live network}
\begin{figure}[t]
  \centering
  \fbox{\includegraphics[width=2.5 in]{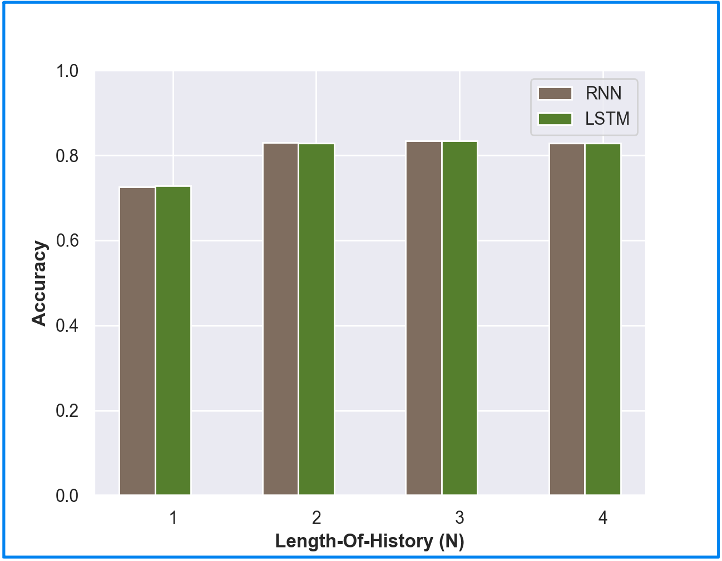}}
  \caption{Accuracy of the beam predictions using the live mmWave network data.}
  \label{fig:ra-1}
\end{figure}

In this section, we shift to beam switching problem where we are interested in identifying the best beam to HO within the serving cell.  The data logs captured from a live  5G mmWave network operated by Ericsson in Chicago is used for the beam prediction problem described in Fig.~\ref{fig:UE_mobility}(b). The commonly used trajectories of UEs defined as evolution in the beam-space of the serving-cell are mined and used to train the RNN structure shown in Fig.~\ref{fig:RNN}. The feature and prediction variables are as given in the equation below: 
\begin{equation}
  \label{eq:ry-1-A}
  \x{i}=\{b_i\} , b_i\in \{1,\ldots,P\},   i\in\{1,\ldots,N\},  
\end{equation}
and
\begin{equation}
  \label{eq:ry-1-B}
  \y{j}=\{b_j\} ,  b_j\in{\rm I\!R}, j\in\{1,\ldots,K\},  
\end{equation}
We use same simulation parameters as shown in the Table~\ref{tab:nn-param}, except that in RNN formulation now we have number of features is equal to number of beams in the serving cell. There are $68$ available beams in the considered serving cell, hence we have $L=P=68$. The performance of predicting the next beam for handover ($K=1$), using the proposed method is given in Fig.~\ref{fig:ra-1}. Notice that as expected, the performance improves with the increase of  $N$, but saturates for $N>2$ indicating that longer historical data does not add much information to the estimation process in practical deployments.

\begin{figure}[t]
  \centering
  \fbox{\includegraphics[width=2.5 in]{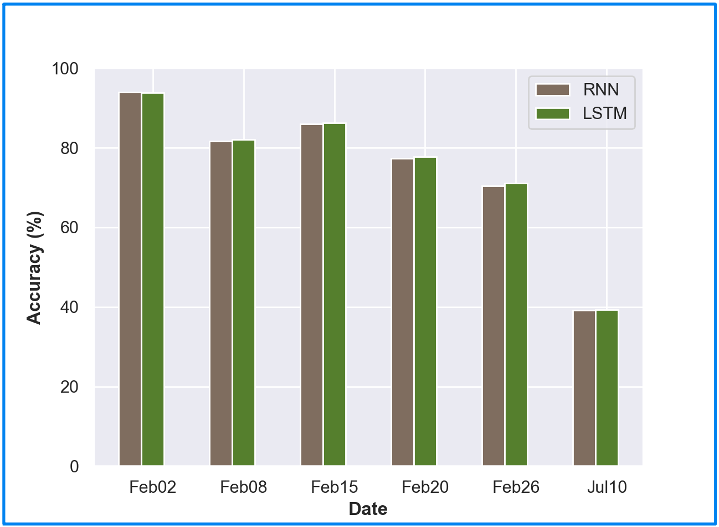}}
  \caption{Accuracy of the beam predictions using the RNN agent trained from the data captured on Feb-01, 2021.}
  \label{fig:ra-2}
\end{figure}

To understand how fast the  trained models degenerate with time, we trained the model with the traces captured on Feb-1st, 2021 and assessed the  performance  on a later date traces. This is shown in Fig.~\ref{fig:ra-2}. Results indicates that the performance degenerate slowly indicating that the AI agent does not require frequent training.

\section{Conclusion and Discussion}
\label{sec:conclusion}
Dense deployment having small cells coupled with high mobility requirements of the 5G and beyond networks will lead to  frequent handover or beam switching thus requiring efficient  HO strategy and beam prediction process. The state of the art handover procedures are reactive in its approach and are sub-optimal for future networks. In this paper, we propose sequence-to-sequence encoding based AI method for mobility prediction, Two specific cases are considered, one for intra-frequency cellular HO and another for beam switching,  depending on the case, the trajectory of the UE is defined in terms of cell-ID/dwell-time or beam-ID sequences.

Results from Fig.~\ref{fig:rx-1} indicate that an accuracy of around $85\%$ can be obtained by exploiting the  historical cell-ID information alone, this can be further improved to around $90\%$ by exploiting the dwell-time in each of the cells as shown in Fig.~\ref{fig:rz-1}. By increasing the value of $K$, there is a slight loss of performance as we are trying to predict far into the future with same amount of historical information. Due to the complex and frequent nature of the handovers in the future cellular network, it would be advantageous to know the sequence of handover BS and when a particular BS is needed to serve the UE beforehand. This can aid in provisioning the required resources in the future HO BSs a-priori. The overall process does not require any measurement from UE's and involves less signaling and latency. 

For the beam switching problem, the proposed method is validated with the live mmWave network in an urban area of Chicago. The results indicate that the performance saturates quickly and there is no significant benefit by having large memory ($N>2$) in the RNN model. Also we notice that the that the performance degenerate slowly and does not require frequent training of the RNN models.
\bibliography{main}
\bibliographystyle{IEEE}
 
\end{document}